\documentclass[8.5pt,twoside,twocolumn]{article}
\usepackage[super,sort&compress,comma]{natbib} 
\usepackage{mhchem}
\usepackage{times,mathptmx}
\usepackage{sectsty}
\usepackage{balance} 

\usepackage{graphicx} 
\usepackage{lastpage}
\usepackage[format=plain,justification=raggedright,singlelinecheck=false,font=small,labelfont=bf,labelsep=space]{caption} 
\usepackage{fancyhdr}
\usepackage{amssymb}

\begin{document}

\thispagestyle{plain}
\fancypagestyle{plain}{
\renewcommand{\headrulewidth}{1pt}}
\renewcommand{\thefootnote}{\fnsymbol{footnote}}
\renewcommand\footnoterule{\vspace*{1pt}%
\hrule width 3.4in height 0.4pt \vspace*{5pt}} 
\setcounter{secnumdepth}{5}

\makeatletter 
\def\subsubsection{\@startsection{subsubsection}{3}{10pt}{-1.25ex plus -1ex minus -.1ex}{0ex plus 0ex}{\normalsize\bf}} 
\def\paragraph{\@startsection{paragraph}{4}{10pt}{-1.25ex plus -1ex minus -.1ex}{0ex plus 0ex}{\normalsize\textit}} 
\renewcommand\@biblabel[1]{#1}            
\renewcommand\@makefntext[1]%
{\noindent\makebox[0pt][r]{\@thefnmark\,}#1}
\makeatother 
\renewcommand{\figurename}{\small{Fig.}~}
\sectionfont{\large}
\subsectionfont{\normalsize} 

\fancyfoot{}
\fancyfoot[LE]{\footnotesize{\sffamily{\thepage~\textbar\hspace{3.45cm} 1--\pageref{LastPage}}}}
\fancyhead{}
\renewcommand{\headrulewidth}{1pt} 
\renewcommand{\footrulewidth}{1pt}
\setlength{\arrayrulewidth}{1pt}
\setlength{\columnsep}{6.5mm}
\setlength\bibsep{1pt}

\twocolumn[
  \begin{@twocolumnfalse}
\noindent\LARGE{\textbf{Structural and microscopic relaxations in a colloidal glass}}
\vspace{0.6cm}

\noindent\large{\textbf{Flavio Augusto de Melo Marques$^{\ast}$\textit{$^{a}$}, Roberta
Angelini \textit{$^{b,c}$}, Emanuela Zaccarelli{$^{c,d}$}, Bela Farago{$^{e}$}, Beatrice Ruta{$^{f}$}, Giancarlo Ruocco\textit{$^{a,c}$}, Barbara Ruzicka\textit{$^{b,c}$} }}\vspace{0.5cm}


\end{@twocolumnfalse}\vspace{0.6cm}
  ]

\noindent\textbf{The aging dynamics of a colloidal glass has been studied by
multiangle Dynamic Light Scattering, Neutron Spin Echo,  X-ray
Photon Correlation Spectroscopy and Molecular Dynamics
simulations. The two relaxation processes, microscopic (fast) and structural (slow),
have been investigated in an unprecedentedly wide range of time
and length scales covering both ergodic and nonergodic regimes.
The microscopic relaxation time remains diffusive at all length scales across the glass
transition scaling with wavevector $Q$ as $Q^{-2}$. The length-scale
dependence of structural relaxation time changes from diffusive, characterized by a $Q^{-2}$-dependence in the early stages of aging, to $Q^{-1}$-dependence in the full aging regime which marks a discontinuous hopping dynamics. Both regimes are associated with a stretched behaviour of the correlation functions. We expect these findings to provide a general
description of both relaxations across the glass transition.}
\section*{}
\vspace{-1cm}


\footnotetext{\textit{$^{a}$~Center for Life Nano Science, IIT@Sapienza,
Istituto Italiano di Tecnologia, Viale Regina Elena 291, 00161
Roma, Italy. Tel: +39 06 4991 3505; E-mail: flavio.marques@iit.it}}
\footnotetext{\textit{$^{b}$~IPCF-CNR, I-00185 Roma, Italy.}}
\footnotetext{\textit{$^{c}$~Dipartimento di Fisica, Sapienza Universit$\grave{a}$ di Roma,
I-00185, Italy.}}
\footnotetext{\textit{$^{d}$~ISC-CNR, I-00185 Roma, Italy.}}
\footnotetext{\textit{$^{e}$~Institute Laue-Langevin, BP 156, 38042 Grenoble, France.}}
\footnotetext{\textit{$^{f}$~European Synchrotron Radiation Facility, B.P. 220 F-38043
Grenoble, Cedex France.}}

\section{Introduction}

The dynamical behaviour of glass-forming
systems has long been the subject of intense
research~\cite{DebenedettiNature2001}. One important feature of
disordered systems
 is that the dynamics of density fluctuations
is characterized by a two-step decay, which implies the presence
of two main relaxation processes~\cite{Balucani}: a microscopic or
fast relaxation, associated to the interactions between a particle
and the cage of its nearest neighbors followed by a structural or
slow relaxation process, related to the structural rearrangements
of the particles. These two distinct processes have been observed
in simple monatomic liquids~\cite{ScopignoPRL2000}, hydrogen
bonded liquids~\cite{AngeliniPRL2002, MonacoPRE1999}, structural
glasses~\cite{MezeiPRL1987, SidebottomPRL1993, ZuriagaPRL2009},
colloids~\cite{RenPRL1993, VanMegenPRL1993, Abou_PRE_2001} and DNA
nanostars~\cite{BiffiPNAS2013}. Furthermore Mode Coupling Theory
(MCT)~\cite{Goetze} has predicted the evolution of the two
relaxations when a system goes from the liquid to the arrested
state: this may happen e.g. by changing temperature, packing
fraction or waiting time (aging). However, even if the supercooled
liquid phase above $T_g$ in glass forming materials has been
widely investigated with both DLS~\cite{SidebottomPRL1993,
SidebottomPRB2007} and neutron scattering
techniques~\cite{MezeiPhysScripta1987, TolleRepProgPhys2001,
ArbePRL2002, FaragoPRE2002, ColmeneroJCP2013}, it is still not
clear how the two relaxation processes behave at the liquid-glass
transition. While in a liquid the dynamics is known to be
diffusive, a change in particle dynamics should occur, as found in
numerical simulations~\cite{AngelaniPRL2000, BroderixPRL2000},
from diffusive to activated in the glass. The signature of this
change at different length scales can be revealed by investigating
the wavevector $Q$ dependence of the relaxation times. Several
numerical studies on different systems such as
water~\cite{SciPRE1996}, ortho-terphenyl~\cite{RinaldiPRE2001} and
hard spheres~\cite{SaltzmanPRE2006} reported a subquadratic
dependence of the structural relaxation time approaching the glass
state. A recent theoretical work based on activated MCT has
rationalised these findings providing a microscopic explanation
for the change of dynamics across the glass
transition~\cite{BhattacharyyaJCP2010}. So far, there has been no
clear experimental proof of this scenario due to technical
difficulties. Moreover, a full description encompassing both
relaxation processes and their behaviour at different length
scales is missing. This is the aim of the present work where the
aging investigation of a colloidal system makes possible a
detailed study across the glass transition, in particular
increasing waiting time in this system plays the same role as
decreasing temperature in glass forming systems~\cite{SciSM2009}.

In this paper we study the aging dynamics of a colloidal glass,
monitoring the waiting time $t_w$ and wavenumber $Q$ dependence of
both fast $\tau_1(Q)$ and slow $\tau_2(Q)$ relaxation times
through a combination of multiangle Dynamic Light Scattering
(DLS), Neutron Spin Echo (NSE), X-Ray Photon Correlation
Spectroscopy (XPCS) and Molecular Dynamics (MD) simulations. In
this way we access an unprecedentedly wide range of time and
length scales and we find a different behaviour for the two
relaxations across the glass transition. While the microscopic one
remains unperturbed, indicating an unchanged, diffusive single
particle dynamics at short times, the structural relaxation shows
a clear change. Indeed, $\tau_2$ going from the liquid to the
arrested state undergoes a transition from a $Q^{-2}$ to a
$Q^{-1}$ behaviour. Pioneering works by Bonn and
coworkers~\cite{Bonn_EL_1999, Abou_PRE_2001} in aqueous Laponite
dispersions reported a $Q^{-2}$ dependence for both microscopic
and structural relaxation times in the ergodic DLS regime. Later
on works on different systems e.g.
 colloids~\cite{Cipelletti_PRL_2000,DuriPRL2009},
clays~\cite{Bellour_PRE_2003, Bandyopadhyay_PRL_2004,
KalounPRE2005,Schoesseler_PRE_2006}, metallic
glasses~\cite{Ruta_PRL_2012},
polymers~\cite{FalusPRL2006,NarayananPRL2007,GuoPRL2009},
supercooled liquids~\cite{CaronnaPRL2008} reported a $Q^{-1}$
dependence of the structural relaxation time associated to an
anomalous compressed exponential relaxation of the correlation
functions attributed to a hyperdiffusive
dynamics~\cite{Bouchaud_EPJE_2001}. Differently, in the present
work
 the relaxation curves at long times are always
described by a stretched exponential, which allow us to interpret
the crossover from a $Q^{-2}$ to a $Q^{-1}$ across the glass
transition as a signature of a change from diffusive dynamics to
discontinuous hopping of caged particles, as predicted
in~\cite{BhattacharyyaJCP2010}.

\section{Materials and Methods}

We used a widely studied colloidal
clay~\cite{Mourchid_Lang_1998,
Mongondry_JCIS_2005,Tanaka_PRE_2005, Jabbari_PRL_2007,
CumminsJNCS2007,Shahin_Langmuir_2010, RuzickaSM2011}, Laponite RD
dispersions, prepared using the same protocol described in
Ref.~\cite{RuzickaSM2011}. All measurements have been performed
using $D_2O$ (EURISO-TOP) of purity $\geq$ 99.9\% as a solvent. As
recently shown the $H/D$ isotopic substitution, required to gain
contrast in neutron scattering measurements, does not
qualitatively affect
 the aging behaviour of
Laponite~\cite{TudiscaRSC2012}. The waiting time origin ($t_w=0$)
is the time at which the dispersion is filtered (and sealed)
directly in glass tubes with diameter of 10 mm for DLS and of 2 mm
for XPCS and in quartz cells with dimensions of 30 mm $\times$ 40
mm $\times$ 4 mm for NSE measurements. All experiments were
performed at the same molar concentration of a $C_w$ = 3.0 \%
sample in salt free water. At this weight concentration  the
system forms a Wigner glass~\cite{Bonn_EL_1999, RuzickaPRL2010}
due to repulsive electrostatic interactions. This is a glass
occurring in a dilute system which shares the main features of
denser glasses, including a two-step
decay~\cite{BeckJCP1999,ZacPRL2008, KangPRE2013}.

DLS measurements were performed with a multi-angle setup in the
time range between $10^{-6}$ s and 1 s. A solid state laser with
wavelength of 642 nm and power of 100 mW and single mode
collecting fibers at five different scattering angles are used.
Time autocorrelation functions are therefore simultaneously
acquired at wavenumbers $Q = 6.2 \times 10^{-4}$, $1.1 \times
10^{-3}$, $1.5 \times 10^{-3}$, $1.8 \times 10^{-3}$ and $2.1
\times 10^{-3}$ \AA$^{-1}$ by calculating the intensity
autocorrelation function as $g_{2}(Q,t)= \frac{\langle
I(Q,t_0)I(Q,t_0+t)\rangle}{\langle I(Q,t_0)\rangle^{2}}$ where
$\langle \cdots \rangle$ denotes the temporal average over $t_0$.
DLS are used only when the system is ergodic in the early aging,
also referred as cage forming, regime characterized by a
structural relaxation time with a waiting time exponential
dependence $\tau_2(t_w) \propto e^{t_w}$~\cite{Tanaka_PRE_2005}.

XPCS measurements were performed at the ID10 beamline  at the
European Synchrotron Radiation Facility (ESRF, Grenoble, France)
in a $Q$-range between  $3.1 \times 10^{-3}$ and $2.2 \times
10^{-2}$ \AA$^{-1}$ including the peak of the static structure
factor occurring at $Q_{peak}\sim 1.6 \times 10^{-2}$ \AA$^{-1}$
as in $H_2O$ solvent~\cite{RuzickaPRL2010}. Using an incident
partially coherent X-ray beam with energy fixed at 8 keV a series
of scattering images were recorded by a charged coupled device
(CCD) and the ensemble averaged intensity autocorrelation function
$g_2(Q, t)$ was calculated by using a standard  multi tau
algorithm after having ensemble averaged over the detector pixels
mapping onto a single Q value~\cite{MadsenNJP2010}. XPCS can thus
be used when the system is non ergodic in the full aging regime
characterized by a waiting time power law dependence of the
structural relaxation time $\tau_2(t_w) \propto t_w^{\alpha}$ with
 $\alpha \sim$ 1~\cite{Tanaka_PRE_2005}.


NSE measurements were performed at the spectrometer IN15 of the
Institute Laue-Langevin (ILL, Grenoble, France) at larger
wavevectors ($1.3 \times 10^{-2} < Q < 1.3 \times 10^{-1}$
\AA$^{-1}$) and at shorter relaxation times (up to $2\times
10^{-7}$ s) with respect to DLS and XPCS. We used wavelength of
10, 16 and 22.8  \AA$^{ }$ yielding time ranges (0.35 - 50) ns, (1.4
- 206) ns and (4.1 - 598) ns respectively. While DLS and XPCS
measure the normalized intensity  correlation function $g_2(Q,t)=
1 + b\left[g_1(Q, t)\right]^2$ (Siegert relation), respectively
through temporal averages (ergodic regime) and ensemble averages
(non ergodic regime), NSE directly accesses the intermediate
scattering function $g_1(Q,t)^2$.

We complement the experimental measurements with MD simulations
for a simple model of low-density glass-former, i.e. a $50-50$
non-crystallising binary mixture of $N=1000$ Yukawa particles of
equal screening length $\xi$ and different repulsion
strength~\cite{ZacPRL2008}\footnote{Lengths and times are measured
in units of $\xi$ and $\xi/(m\epsilon)^{1/2}$ respectively, where
$m$ is the mass of the particles and $\epsilon$ is the unit of
energy. Temperature is measured in units of $\epsilon$ (i.e.
$k_B$=1 where $k_B$ is the Boltzmann constant).}. The system was
found~\cite{ZacPRL2008} to undergo a Wigner glass transition upon
decreasing temperature $T$. To mimic the experimental situation,
we performed a quench inside the glassy region at fixed number
density $\rho=0.002984$. The system was equilibrated at high
$T=1.0\times10^{-3}$ and then instantaneously quenched to
$T=1.6\times10^{-5}$, below the glass transition occurring at
$T_g\sim 1.7\times10^{-5}$. The waiting time origin was the time
of the quench. Self intermediate scattering functions $F^s(Q,t)$
have been calculated for different wave vectors as a function of
waiting time, averaging over 20 independent quenches to improve
statistics.
\begin{figure}[t]
\centering
\includegraphics[width=7cm]{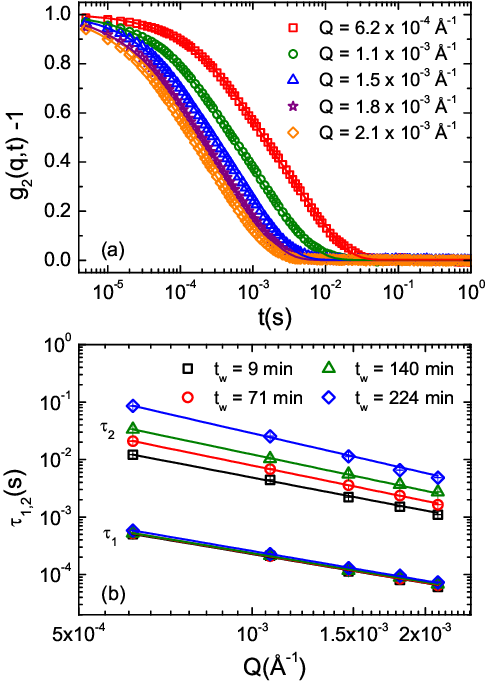} \\
\caption{(a) Normalized intensity correlation functions measured
through DLS for different $Q$-values at $t_w = 9$ min. Solid lines
are the fits obtained through Eq. \ref{eqfitDLS} with $\gamma$=1.
(b) $Q$-dependence of the fast $\tau_1$ and slow $\tau_2$
relaxation times at different waiting times. Solid lines are power
law fits of the data as $\sim Q^{-n}$ where $n$=1.70 $\pm$ 0.06
for $\tau_1$ and $n$=2.02 $\pm$ 0.19 for
$\tau_2$.}\label{figureDLS}
\end{figure}

\section{Results}

Figure~\ref{figureDLS}(a) shows the normalized
intensity correlation functions measured by  DLS at initial
waiting time ($t_w = 9$ min) for different $Q$-values (symbols)
and the corresponding fits (full lines) obtained through a typical
double exponential decay:
\begin{equation}\label{eqfitDLS}
g_{2}(q,t)-1=b\left[a
e^{-\big(\frac{t}{\tau_{1}}\big)^{\gamma}}+(1-a)
e^{-\big(\frac{t}{\tau_{2}}\big)^{\beta}}\right]^{2}
\end{equation}
\noindent where the parameters $a$ and $(1-a)$ are the amplitudes
of the two relaxation modes, $b$ is the coherence factor, $\tau_1$
is
 the fast relaxation time connected to the microscopic motion of particles, $\tau_2$
is the slow relaxation time related to the structural
rearrangement, $\gamma$ (here $\gamma$=1~\cite{Abou_PRE_2001,
RuzickaPRL2004}) and $\beta$ measure the distribution widths of
the two relaxations.

The $Q$-dependence  of the fast and slow relaxation times is
reported in Fig. \ref{figureDLS}(b) at different waiting times.
While $\tau_1$ shows a moderate waiting time dependence, $\tau_2$
increases significantly with $t_w$. Both times are well described
by power law fits $\sim Q^{-n}$ with $n\sim 2$ (solid lines in
Fig. \ref{figureDLS}(b)). Hence both microscopic and structural
relaxation times display an almost quadratic wave vector
dependence in the DLS (early aging) regime, which is the signature
of diffusive dynamics for both relaxations at these early waiting
times. These findings are in agreement with the works on Laponite
water suspensions by the group of Bonn et al.~\cite{Bonn_EL_1999,
Abou_PRE_2001} and of Munch et al.~\cite{KalounPRE2005,
Schoesseler_PRE_2006}.
\begin{figure}[t]
\centering
\includegraphics[width=6.8cm]{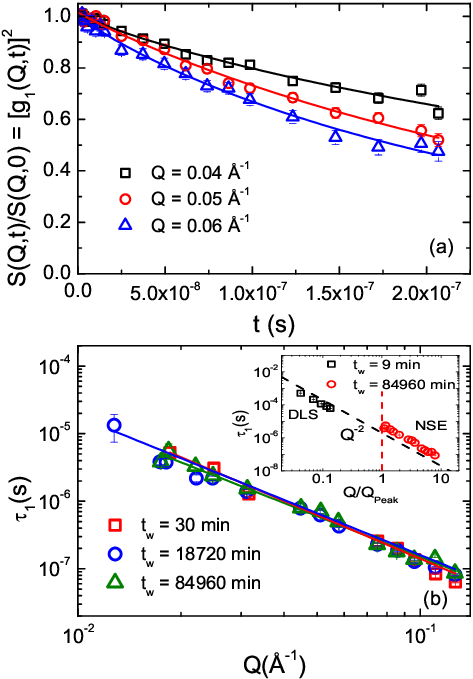} \\
\caption{(a) Dynamic structure factor obtained from NSE
measurements for different $Q$-values at $t_w$ = 30 min. Solid
lines are stretched exponential fits with $\beta \sim$ 1.(b)
$Q$-dependence of the fast relaxation time at different waiting
times. Solid lines are power law fits of the data as $\sim Q^{-n}$
where $n$=1.97 $\pm$ 0.24. Inset: comparison between fast
relaxation times measured through DLS (t$_w$=9 min) and NSE
(t$_w$=84960 min) as a function of Q/Q$_{peak}$. The dashed line
indicates the $Q^{-2}$ behaviour. The red vertical line indicates
the position of the structure factor peak.}\label{figureNSE}
\end{figure}
Fig.~\ref{figureNSE}(a) shows the dynamic structure factors
measured by NSE at different wavevectors (symbols) together
with single stretched exponential fits (full lines). The corresponding
 fast relaxation times are  shown in Fig.~\ref{figureNSE}(b). We find
 that $\tau_1$  scales as $ \sim
Q^{-2}$ during the whole experiment. As reported in the inset of
Fig.~\ref{figureNSE}(b) the estimate of $\tau_1$ obtained from the
NSE fits is in good agreement with that obtained by DLS, taking
into account the large difference in $Q$ between these techniques.
Since the studied sample experiences a sol to Wigner glass
transition at $t_w \approx$ 600 min, NSE results ensure that the
microscopic relaxation time remarkably scales as $Q^{-2}$ both in
the early aging and full aging regimes indicating that the
short-time dynamics remains diffusive for all the investigated
dynamical range even in the arrested state.

\begin{figure}[t]
\centering
\includegraphics[width=7.7cm]{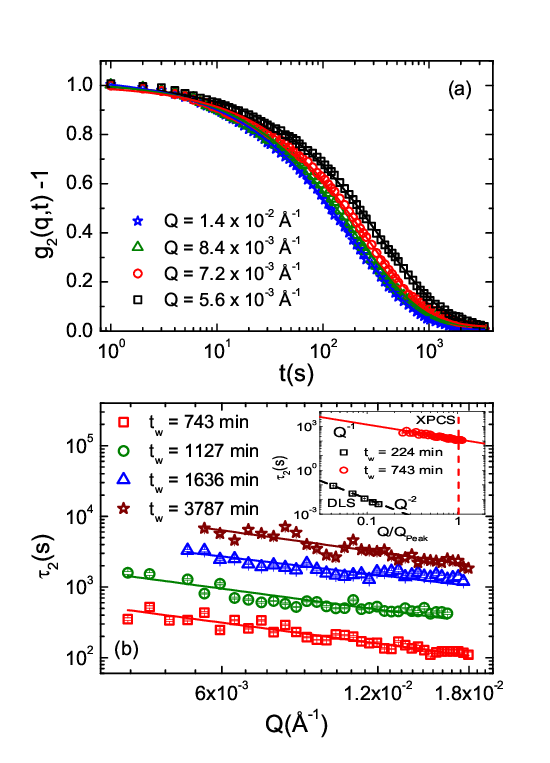}
\caption{(a) Normalized intensity correlation functions measured
through XPCS for different $Q$-values at $t_w = 1127$ min. Solid
lines are the fits obtained through the stretched exponential
decay. (b) $Q$-dependence of the slow relaxation time
 at different waiting times. Solid lines are power law fits of the data as $\sim Q^{-n}$
where $n$=0.95 $\pm$ 0.12. Inset: comparison between slow
relaxation times measured through DLS (t$_w$=9 min) and XPCS
(t$_w$=743 min) as a function of Q/Q$_{peak}$. The dashed and full
lines indicate respectively the $Q^{-2}$ and $Q^{-1}$ behaviours.
The red vertical line indicates the position of the structure
factor peak.}\label{figureXPCS}
\end{figure}

In Fig. \ref{figureXPCS}(a) normalized intensity correlation
functions probed by XPCS in the full aging regime are shown at
different $Q$-values  (symbols). At these long waiting times the
fast relaxation time $\tau_1$ is out of the XPCS detection window.
Hence only the slow relaxation time $\tau_2$ can be measured and
 only the second term of Eq. \ref{eqfitDLS} is used to
fit the data (full lines in Fig. \ref{figureXPCS}(a)). We notice
that in our case DLS measurements allow to probe relaxation times
up to $\sim$10$^{-1}$ s (higher limit in Fig.~\ref{figureDLS} (b)),
reached for the sample in the ergodic region at waiting time
$t_w$=224 min,
while XPCS measurements permit to access relaxation times above
$\sim$10$^2$  s (lower limit in Fig 3b) achieved for the sample in
the non ergodic region at waiting times $t_w$=743 min. Figure
\ref{figureXPCS}(b) shows the $Q$-dependence of $\tau_2$ at
different waiting times (symbols) and the associated power law
fits (full lines) as $\sim Q^{-n}$. In this regime we find $n
\simeq$ 1 ruling out diffusive dynamics. Therefore a crossover
from the DLS early aging regime characterized by $\tau_2(Q)
\propto Q^{-2}$, $\tau_2(t_w) \propto e^{t_w}$ to the XPCS full
aging regime characterized by $\tau_2(Q) \propto Q^{-1}$,
$\tau_2(t_w) \propto t_w^{\alpha}$ with $\alpha \sim$ 1 is observed. In both cases the intensity
correlation functions are described by stretched exponentials at
variance with the $\beta>$1 behaviour found in previous
works~\cite{Bellour_PRE_2003, Bandyopadhyay_PRL_2004,
KalounPRE2005, Schoesseler_PRE_2006} and in rejuvenated
samples~\cite{AngeliniSM2013}, as fully discussed in
Ref.~\cite{AngeliniSM2013}.

The combination of three complementary experimental techniques has
allowed us to show that the fast relaxation mode remains diffusive
both in ergodic and non-ergodic conditions at all waiting times.
On the other hand, we observe that the structural relaxation time
is characterized by two distinct $Q$ behaviours in the two
different $t_w$ regimes. It remains to elucidate what happens in
between the DLS and XPCS regimes, i.e. whether the change is
discontinuous or not. Bhattacharyya et
al.~\cite{BhattacharyyaJCP2010} have predicted a gradual change of
the dynamics. To address this point we turn to MD simulations.

\begin{figure}[t]
\centering
\includegraphics[width=7cm]{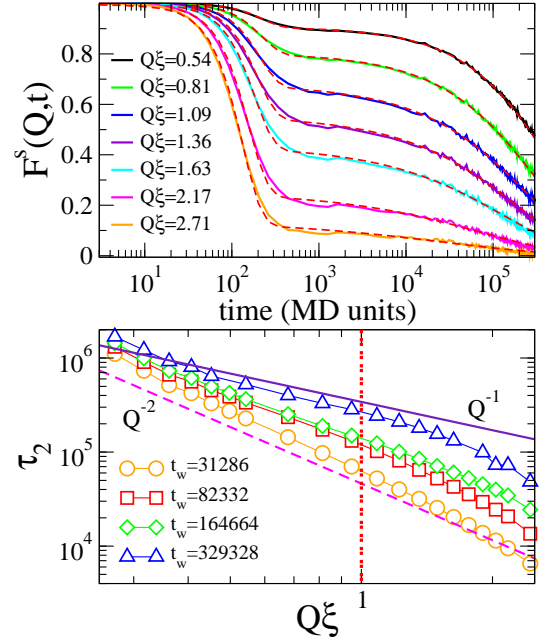}
\caption{(a) Self intermediate scattering functions (full lines)
from MD simulations following a quench to $T=1.6\times 10^{-5}$ at
fixed waiting time $t_w=329328$ MD units and different wave
vectors, as reported in the labels. Dashed lines are the fitting
curves obtained using the expression within the square brackets of
Eq.~\ref{eqfitDLS} with $\gamma$=2. (b) $Q$-dependence of the slow
relaxation time $\tau_2$ extracted from the fits for selected
values of $t_w$. The dashed and full lines indicate respectively
the $Q^{-2}$ and $Q^{-1}$ behaviours. The red vertical line
indicates the position of the structure factor peak. }
\label{figureMD-Q}
\end{figure}

In Fig.~\ref{figureMD-Q}(a) we report the MD self intermediate
scattering functions (full lines) at fixed waiting time after a
quench in the glassy state for several wavevectors. The
microscopic dynamics is Newtonian, thereby the data are fitted by
the double exponential decay within the square brackets of
Eq.~\ref{eqfitDLS} fixing $\gamma$=2 and the corresponding
$\tau_1$ is not relevant to describe the experimental (Brownian)
fast relaxation. We discuss only wavevectors small enough that the
second exponential in Eq.~\ref{eqfitDLS} is stretched ($\beta <
1$), excluding large $Q$ values where the dynamics becomes again
ballistic due to the underlying microscopic
dynamics~\cite{SawPRL2009}. In the early aging regime, the
correlators do not display the typical two-step decay, similarly
to what observed in out-of-equilibrium simulations of other glass
formers~\cite{MasriPRE2010}. We thus focus on the regime occurring
at larger waiting times, i.e. $t_w \gtrsim 20000$, where full aging
starts and the double exponential describes the correlators, as
shown by the fits in Fig.~\ref{figureMD-Q}(a) (dashed lines). The
corresponding $\tau_2$ is reported as a function of $Q$ in
Fig.~\ref{figureMD-Q}(b) for different values of the waiting time.
We find that $\tau_2$ initially displays a $Q^{-2}$ dependence at
low and intermediate wave-vectors, in agreement with the DLS data
of Fig.~\ref{figureDLS}. Upon increasing $t_w$, while at low $Q$ (approaching the
hydrodynamic regime) $\tau_2$ still tends to recover a diffusive
behaviour ($Q^{-2}$), at intermediate $Q$ encompassing the main
structure factor peak  (see vertical line in
Fig.~\ref{figureMD-Q}b) a clear power-law dependence $Q^{-n}$ with
the exponent lowering and approaching unity at large $t_w$, is
observed. At even larger $t_w$ the system becomes non-ergodic on
the timescale of our simulations. Thus, in the full aging regime
we observe a strongly non-diffusive dependence of $\tau_2$,
approaching a $Q^{-1}$ behaviour, in full agreement with the XPCS
data in the same $Q$-window. Interestingly, $n$ decreases
continuously with increasing $t_w$ in agreement with theoretical
predictions~\cite{BhattacharyyaJCP2010}.

\section{Conclusions}

In this work we have investigated the
$Q$-dependence of both microscopic and structural relaxation times
during the early aging and full aging regimes of a colloidal glass
through multiangle DLS, NSE and XPCS techniques, covering a wide
range of wavevectors and times. The experimental results have been
complemented by MD simulations. We found that the microscopic
relaxation time, $\tau_1$, characteristic of the short-time
diffusion of a particle in the suspending medium, scales as
$Q^{-2}$ during both early aging and full aging regimes, depicting
a diffusive nature of particles motion. On the contrary the slow
relaxation time, $\tau_2$, associated to the structural
rearrangement of the system, shows a transition from a $Q^{-2}$
diffusive behaviour in the liquid (or early aging) regime
 to a $Q^{-1}$ activated dynamics in the glass (or full aging)
regime associated with a stretched behaviour of the correlation functions, in full agreement with recent theoretical
predictions~\cite{BhattacharyyaJCP2010}. The reported experimental
evidence is not specific to the studied system, but appears to
have an intrinsic generality,
 occurring also in numerical studies on different glass-formers~\cite{SciPRE1996,RinaldiPRE2001,SaltzmanPRE2006}.
Therefore our study of both relaxations in an aging colloidal
system may provide a general description of the complex dynamics
at fast and slow timescales of any glass-former.

\section{Acknowledgments} 

We acknowledge ILL and ESRF for beamtime
and Orsolya Czakkel for providing assistance during the XPCS
measurements. RA and EZ acknowledge support from MIUR-PRIN. EZ
acknowledges support from MIUR-FIRB ANISOFT (RBFR125H0M).

\footnotesize{\bibliographystyle{rsc}

\begin{thebibliography}{56}
\expandafter\ifx\csname
natexlab\endcsname\relax\def\natexlab#1{#1}\fi
\expandafter\ifx\csname bibnamefont\endcsname\relax
  \def\bibnamefont#1{#1}\fi
\expandafter\ifx\csname bibfnamefont\endcsname\relax
  \def\bibfnamefont#1{#1}\fi
\expandafter\ifx\csname citenamefont\endcsname\relax
  \def\citenamefont#1{#1}\fi
\expandafter\ifx\csname url\endcsname\relax
  \def\url#1{\texttt{#1}}\fi
\expandafter\ifx\csname
urlprefix\endcsname\relax\def\urlprefix{URL }\fi
\providecommand{\bibinfo}[2]{#2}
\providecommand{\eprint}[2][]{\url{#2}}

\bibitem[{\citenamefont{Debenedetti and
  Stillinger}(2001)}]{DebenedettiNature2001}
\bibinfo{author}{\bibfnamefont{P.~G.} \bibnamefont{Debenedetti}}
  \bibnamefont{and} \bibinfo{author}{\bibfnamefont{F.~H.}
  \bibnamefont{Stillinger}}, \bibinfo{journal}{Nature}
  \textbf{\bibinfo{volume}{410}}, \bibinfo{pages}{259} (\bibinfo{year}{2001}).

\bibitem[{\citenamefont{Balucani and Zoppi}(1994)}]{Balucani}
\bibinfo{author}{\bibfnamefont{U.}~\bibnamefont{Balucani}} \bibnamefont{and}
  \bibinfo{author}{\bibfnamefont{M.}~\bibnamefont{Zoppi}},
  \emph{\bibinfo{title}{Dynamics of the Liquid State}}
  (\bibinfo{publisher}{Clarendon Press - Oxford}, \bibinfo{year}{1994}).

\bibitem[{\citenamefont{Scopigno et~al.}(2000)\citenamefont{Scopigno, Balucani,
  Ruocco, and Sette}}]{ScopignoPRL2000}
\bibinfo{author}{\bibfnamefont{T.}~\bibnamefont{Scopigno}},
  \bibinfo{author}{\bibfnamefont{U.}~\bibnamefont{Balucani}},
  \bibinfo{author}{\bibfnamefont{G.}~\bibnamefont{Ruocco}}, \bibnamefont{and}
  \bibinfo{author}{\bibfnamefont{F.}~\bibnamefont{Sette}},
  \bibinfo{journal}{Phys. Rev. Lett.} \textbf{\bibinfo{volume}{85}},
  \bibinfo{pages}{4076} (\bibinfo{year}{2000}).

\bibitem[{\citenamefont{Angelini et~al.}(2002)\citenamefont{Angelini, Giura,
  Monaco, Ruocco, Sette, and Verbeni}}]{AngeliniPRL2002}
\bibinfo{author}{\bibfnamefont{R.}~\bibnamefont{Angelini}},
  \bibinfo{author}{\bibfnamefont{P.}~\bibnamefont{Giura}},
  \bibinfo{author}{\bibfnamefont{G.}~\bibnamefont{Monaco}},
  \bibinfo{author}{\bibfnamefont{G.}~\bibnamefont{Ruocco}},
  \bibinfo{author}{\bibfnamefont{F.}~\bibnamefont{Sette}}, \bibnamefont{and}
  \bibinfo{author}{\bibfnamefont{R.}~\bibnamefont{Verbeni}},
  \bibinfo{journal}{Phys. Rev. Lett.} \textbf{\bibinfo{volume}{88}},
  \bibinfo{pages}{255503} (\bibinfo{year}{2002}).

\bibitem[{\citenamefont{Monaco et~al.}(1999)\citenamefont{Monaco, Cunsolo,
  Ruocco, and Sette}}]{MonacoPRE1999}
\bibinfo{author}{\bibfnamefont{G.}~\bibnamefont{Monaco}},
  \bibinfo{author}{\bibfnamefont{A.}~\bibnamefont{Cunsolo}},
  \bibinfo{author}{\bibfnamefont{G.}~\bibnamefont{Ruocco}}, \bibnamefont{and}
  \bibinfo{author}{\bibfnamefont{F.}~\bibnamefont{Sette}},
  \bibinfo{journal}{Phys. Rev. E} \textbf{\bibinfo{volume}{60}},
  \bibinfo{pages}{5505} (\bibinfo{year}{1999}).

\bibitem[{\citenamefont{Mezei, Knaak and Farago}(1987{\natexlab{a}})}]{MezeiPRL1987}
  \bibinfo{author}{\bibfnamefont{F.}~\bibnamefont{Mezei}},
  \bibinfo{author}{\bibfnamefont{W.}~\bibnamefont{Knaak}}, \bibnamefont{and}
  \bibinfo{author}{\bibfnamefont{B.}~\bibnamefont{Farago}},
  \bibinfo{journal}{Phys. Rev. Lett.} \textbf{\bibinfo{volume}{58}},
  \bibinfo{pages}{571} (\bibinfo{year}{1987}{\natexlab{a}}).

\bibitem[{\citenamefont{Sidebottom et~al.}(1993)\citenamefont{Sidebottom,
  Bergman, Borjesson, and Torell}}]{SidebottomPRL1993}
\bibinfo{author}{\bibfnamefont{D.~L.} \bibnamefont{Sidebottom}},
  \bibinfo{author}{\bibfnamefont{R.}~\bibnamefont{Bergman}},
  \bibinfo{author}{\bibfnamefont{L.}~\bibnamefont{Borjesson}},
  \bibnamefont{and} \bibinfo{author}{\bibfnamefont{L.~M.}
  \bibnamefont{Torell}}, \bibinfo{journal}{Phys. Rev. Lett.}
  \textbf{\bibinfo{volume}{71}}, \bibinfo{pages}{2260} (\bibinfo{year}{1993}).

\bibitem[{\citenamefont{Zuriaga et~al.}(2009)\citenamefont{Zuriaga, Pardo,
  Lunkenheimer, Tamarit, Veglio, Barrio, Bermejo, and Loidl}}]{ZuriagaPRL2009}
\bibinfo{author}{\bibfnamefont{M.}~\bibnamefont{Zuriaga}},
  \bibinfo{author}{\bibfnamefont{L.~C.} \bibnamefont{Pardo}},
  \bibinfo{author}{\bibfnamefont{P.}~\bibnamefont{Lunkenheimer}},
  \bibinfo{author}{\bibfnamefont{J.~L.} \bibnamefont{Tamarit}},
  \bibinfo{author}{\bibfnamefont{N.}~\bibnamefont{Veglio}},
  \bibinfo{author}{\bibfnamefont{M.}~\bibnamefont{Barrio}},
  \bibinfo{author}{\bibfnamefont{F.~J.} \bibnamefont{Bermejo}},
  \bibnamefont{and} \bibinfo{author}{\bibfnamefont{A.}~\bibnamefont{Loidl}},
  \bibinfo{journal}{Phys. Rev. Lett.} \textbf{\bibinfo{volume}{103}},
  \bibinfo{pages}{075701} (\bibinfo{year}{2009}).

\bibitem[{\citenamefont{Ren and Sorensen}(1993)}]{RenPRL1993}
\bibinfo{author}{\bibfnamefont{S.~Z.} \bibnamefont{Ren}} \bibnamefont{and}
  \bibinfo{author}{\bibfnamefont{C.~M.} \bibnamefont{Sorensen}},
  \bibinfo{journal}{Phys. Rev. Lett.} \textbf{\bibinfo{volume}{70}},
  \bibinfo{pages}{1727} (\bibinfo{year}{1993}).

\bibitem[{\citenamefont{van Megen and Underwood}(1993)}]{VanMegenPRL1993}
\bibinfo{author}{\bibfnamefont{W.}~\bibnamefont{van Megen}} \bibnamefont{and}
  \bibinfo{author}{\bibfnamefont{S.~M.} \bibnamefont{Underwood}},
  \bibinfo{journal}{Physical Review Letters} \textbf{\bibinfo{volume}{70}},
  \bibinfo{pages}{2766} (\bibinfo{year}{1993}).

\bibitem[{\citenamefont{Abou et~al.}(2001)\citenamefont{Abou, Bonn, and
  Meunier}}]{Abou_PRE_2001}
\bibinfo{author}{\bibfnamefont{B.}~\bibnamefont{Abou}},
  \bibinfo{author}{\bibfnamefont{D.}~\bibnamefont{Bonn}}, \bibnamefont{and}
  \bibinfo{author}{\bibfnamefont{J.}~\bibnamefont{Meunier}},
  \bibinfo{journal}{Phys. Rev. E} \textbf{\bibinfo{volume}{64}},
  \bibinfo{pages}{021510} (\bibinfo{year}{2001}).

\bibitem[{\citenamefont{Biffi et~al.}(2013)\citenamefont{Biffi, Cerbino,
  Bomboi, Paraboschia, Asselta, Sciortino, and Bellini}}]{BiffiPNAS2013}
\bibinfo{author}{\bibfnamefont{S.}~\bibnamefont{Biffi}},
  \bibinfo{author}{\bibfnamefont{R.}~\bibnamefont{Cerbino}},
  \bibinfo{author}{\bibfnamefont{F.}~\bibnamefont{Bomboi}},
  \bibinfo{author}{\bibfnamefont{E.}~\bibnamefont{Paraboschia}},
  \bibinfo{author}{\bibfnamefont{R.}~\bibnamefont{Asselta}},
  \bibinfo{author}{\bibfnamefont{F.}~\bibnamefont{Sciortino}},
  \bibnamefont{and} \bibinfo{author}{\bibfnamefont{T.}~\bibnamefont{Bellini}},
  \bibinfo{journal}{PNAS} \textbf{\bibinfo{volume}{24}}, \bibinfo{pages}{15633}
  (\bibinfo{year}{2013}).

\bibitem[{\citenamefont{G{\"o}tze}(2008)}]{Goetze}
\bibinfo{author}{\bibfnamefont{W.}~\bibnamefont{G{\"o}tze}},
  \emph{\bibinfo{title}{Complex Dynamics of Glass-Forming Liquids: A
  Mode-Coupling Theory}} (\bibinfo{publisher}{Oxford Science Publications -
  Oxford}, \bibinfo{year}{2008}).

\bibitem[{\citenamefont{Sidebottom et~al.}(2007)\citenamefont{Sidebottom,
  Rodenburg, and Changstrom}}]{SidebottomPRB2007}
\bibinfo{author}{\bibfnamefont{D.~L.} \bibnamefont{Sidebottom}},
  \bibinfo{author}{\bibfnamefont{B.~V.} \bibnamefont{Rodenburg}},
  \bibnamefont{and} \bibinfo{author}{\bibfnamefont{J.~R.}
  \bibnamefont{Changstrom}}, \bibinfo{journal}{Phys. Rev. B}
  \textbf{\bibinfo{volume}{75}}, \bibinfo{pages}{132201}
  (\bibinfo{year}{2007}).

\bibitem[{\citenamefont{Mezei and
  Knaak}(1987{\natexlab{b}})}]{MezeiPhysScripta1987}
\bibinfo{author}{\bibfnamefont{F.}~\bibnamefont{Mezei}} \bibnamefont{and}
  \bibinfo{author}{\bibfnamefont{W.}~\bibnamefont{Knaak}},
  \bibinfo{journal}{Phys. Scripta} \textbf{\bibinfo{volume}{T19}},
  \bibinfo{pages}{363} (\bibinfo{year}{1987}{\natexlab{b}}).

\bibitem[{\citenamefont{Tolle}(2001)}]{TolleRepProgPhys2001}
\bibinfo{author}{\bibfnamefont{A.}~\bibnamefont{Tolle}}, \bibinfo{journal}{Rep.
  Prog. Phys.} \textbf{\bibinfo{volume}{64}}, \bibinfo{pages}{1473}
  (\bibinfo{year}{2001}).

\bibitem[{\citenamefont{Arbe et~al.}(2002)\citenamefont{Arbe, Colmenero,
  Alvarez, Monkenbusch, Richter, Farago, and Frick}}]{ArbePRL2002}
\bibinfo{author}{\bibfnamefont{A.}~\bibnamefont{Arbe}},
  \bibinfo{author}{\bibfnamefont{J.}~\bibnamefont{Colmenero}},
  \bibinfo{author}{\bibfnamefont{F.}~\bibnamefont{Alvarez}},
  \bibinfo{author}{\bibfnamefont{M.}~\bibnamefont{Monkenbusch}},
  \bibinfo{author}{\bibfnamefont{D.}~\bibnamefont{Richter}},
  \bibinfo{author}{\bibfnamefont{B.}~\bibnamefont{Farago}}, \bibnamefont{and}
  \bibinfo{author}{\bibfnamefont{B.}~\bibnamefont{Frick}},
  \bibinfo{journal}{Phys. Rev. Lett.} \textbf{\bibinfo{volume}{89}},
  \bibinfo{pages}{245701} (\bibinfo{year}{2002}).

\bibitem[{\citenamefont{Farago et~al.}(2002)\citenamefont{Farago, Arbe,
  Colmenero, Faust, Buchenau, and Richter}}]{FaragoPRE2002}
\bibinfo{author}{\bibfnamefont{B.}~\bibnamefont{Farago}},
  \bibinfo{author}{\bibfnamefont{A.}~\bibnamefont{Arbe}},
  \bibinfo{author}{\bibfnamefont{J.}~\bibnamefont{Colmenero}},
  \bibinfo{author}{\bibfnamefont{R.}~\bibnamefont{Faust}},
  \bibinfo{author}{\bibfnamefont{U.}~\bibnamefont{Buchenau}}, \bibnamefont{and}
  \bibinfo{author}{\bibfnamefont{D.}~\bibnamefont{Richter}},
  \bibinfo{journal}{Phys. Rev. E} \textbf{\bibinfo{volume}{65}},
  \bibinfo{pages}{051803} (\bibinfo{year}{2002}).

\bibitem[{\citenamefont{Colmenero et~al.}(2013)\citenamefont{Colmenero,
  Alvarez, Khairy, and Arbe}}]{ColmeneroJCP2013}
\bibinfo{author}{\bibfnamefont{J.}~\bibnamefont{Colmenero}},
  \bibinfo{author}{\bibfnamefont{F.}~\bibnamefont{Alvarez}},
  \bibinfo{author}{\bibfnamefont{Y.}~\bibnamefont{Khairy}}, \bibnamefont{and}
  \bibinfo{author}{\bibfnamefont{A.}~\bibnamefont{Arbe}}, \bibinfo{journal}{J.
  Chem. Phys.} \textbf{\bibinfo{volume}{139}}, \bibinfo{pages}{044906}
  (\bibinfo{year}{2013}).

\bibitem[{\citenamefont{L.Angelani et~al.}(2000)\citenamefont{L.Angelani,
  Leonardo, Ruocco, Scala, and Sciortino}}]{AngelaniPRL2000}
\bibinfo{author}{\bibnamefont{L.Angelani}},
  \bibinfo{author}{\bibfnamefont{R.} \bibnamefont{DiLeonardo}},
  \bibinfo{author}{\bibfnamefont{G.}~\bibnamefont{Ruocco}},
  \bibinfo{author}{\bibfnamefont{A.}~\bibnamefont{Scala}}, \bibnamefont{and}
  \bibinfo{author}{\bibfnamefont{F.}~\bibnamefont{Sciortino}},
  \bibinfo{journal}{Phys. Rev. Lett.} \textbf{\bibinfo{volume}{85}},
  \bibinfo{pages}{5356} (\bibinfo{year}{2000}).

\bibitem[{\citenamefont{Broderix et~al.}(2000)\citenamefont{Broderix,
  Bhattacharya, Cavagna, Zippelius, and Giardina}}]{BroderixPRL2000}
\bibinfo{author}{\bibfnamefont{K.}~\bibnamefont{Broderix}},
  \bibinfo{author}{\bibfnamefont{K.~K.} \bibnamefont{Bhattacharya}},
  \bibinfo{author}{\bibfnamefont{A.}~\bibnamefont{Cavagna}},
  \bibinfo{author}{\bibfnamefont{A.}~\bibnamefont{Zippelius}},
  \bibnamefont{and} \bibinfo{author}{\bibfnamefont{I.}~\bibnamefont{Giardina}},
  \bibinfo{journal}{Phys. Rev. Lett.} \textbf{\bibinfo{volume}{85}},
  \bibinfo{pages}{5360} (\bibinfo{year}{2000}).

\bibitem[{\citenamefont{Sciortino et~al.}(1996)\citenamefont{Sciortino, Gallo,
  Tartaglia, and Chen}}]{SciPRE1996}
\bibinfo{author}{\bibfnamefont{F.}~\bibnamefont{Sciortino}},
  \bibinfo{author}{\bibfnamefont{P.}~\bibnamefont{Gallo}},
  \bibinfo{author}{\bibfnamefont{P.}~\bibnamefont{Tartaglia}},
  \bibnamefont{and} \bibinfo{author}{\bibfnamefont{S.~H.} \bibnamefont{Chen}},
  \bibinfo{journal}{Phys. Rev. E} \textbf{\bibinfo{volume}{54}},
  \bibinfo{pages}{6331} (\bibinfo{year}{1996}).

\bibitem[{\citenamefont{Rinaldi et~al.}(2001)\citenamefont{Rinaldi, Sciortino,
  and Tartaglia}}]{RinaldiPRE2001}
\bibinfo{author}{\bibfnamefont{A.}~\bibnamefont{Rinaldi}},
  \bibinfo{author}{\bibfnamefont{F.}~\bibnamefont{Sciortino}},
  \bibnamefont{and}
  \bibinfo{author}{\bibfnamefont{P.}~\bibnamefont{Tartaglia}},
  \bibinfo{journal}{Phys. Rev. E} \textbf{\bibinfo{volume}{63}},
  \bibinfo{pages}{061210} (\bibinfo{year}{2001}).

\bibitem[{\citenamefont{Saltzman and Schweizer}(2006)}]{SaltzmanPRE2006}
\bibinfo{author}{\bibfnamefont{E.~J.} \bibnamefont{Saltzman}} \bibnamefont{and}
  \bibinfo{author}{\bibfnamefont{K.~S.} \bibnamefont{Schweizer}},
  \bibinfo{journal}{Phys. Rev. E} \textbf{\bibinfo{volume}{74}},
  \bibinfo{pages}{061501} (\bibinfo{year}{2006}).

\bibitem[{\citenamefont{Bhattacharyya et~al.}(2010)\citenamefont{Bhattacharyya,
  Bagchi, and Wolynes}}]{BhattacharyyaJCP2010}
\bibinfo{author}{\bibfnamefont{S.~M.} \bibnamefont{Bhattacharyya}},
  \bibinfo{author}{\bibfnamefont{B.}~\bibnamefont{Bagchi}}, \bibnamefont{and}
  \bibinfo{author}{\bibfnamefont{P.~G.} \bibnamefont{Wolynes}},
  \bibinfo{journal}{J. Chem. Phys.} \textbf{\bibinfo{volume}{132}},
  \bibinfo{pages}{104503} (\bibinfo{year}{2010}).

\bibitem[{\citenamefont{Sciortino et~al.}(2009)\citenamefont{Sciortino,
  Michele, Corezzi, Russo, Zaccarelli, and P.Tartaglia}}]{SciSM2009}
\bibinfo{author}{\bibfnamefont{F.}~\bibnamefont{Sciortino}},
  \bibinfo{author}{\bibfnamefont{C.~D.} \bibnamefont{Michele}},
  \bibinfo{author}{\bibfnamefont{S.}~\bibnamefont{Corezzi}},
  \bibinfo{author}{\bibfnamefont{J.}~\bibnamefont{Russo}},
  \bibinfo{author}{\bibfnamefont{E.}~\bibnamefont{Zaccarelli}},
  \bibnamefont{and} \bibinfo{author}{\bibnamefont{P.Tartaglia}},
  \bibinfo{journal}{Soft Matter} \textbf{\bibinfo{volume}{5}},
  \bibinfo{pages}{2571} (\bibinfo{year}{2009}).

\bibitem[{\citenamefont{Bonn et~al.}(1999)\citenamefont{Bonn, Tanaka, Wegdam,
  Kellay, and Meunier}}]{Bonn_EL_1999}
\bibinfo{author}{\bibfnamefont{D.}~\bibnamefont{Bonn}},
  \bibinfo{author}{\bibfnamefont{H.}~\bibnamefont{Tanaka}},
  \bibinfo{author}{\bibfnamefont{G.}~\bibnamefont{Wegdam}},
  \bibinfo{author}{\bibfnamefont{H.}~\bibnamefont{Kellay}}, \bibnamefont{and}
  \bibinfo{author}{\bibfnamefont{J.}~\bibnamefont{Meunier}},
  \bibinfo{journal}{Europhys. Lett.} \textbf{\bibinfo{volume}{45}},
  \bibinfo{pages}{52} (\bibinfo{year}{1999}).

\bibitem[{\citenamefont{Cipelletti et~al.}(2000)\citenamefont{Cipelletti,
  Manley, Ball, and Weitz}}]{Cipelletti_PRL_2000}
\bibinfo{author}{\bibfnamefont{L.}~\bibnamefont{Cipelletti}},
  \bibinfo{author}{\bibfnamefont{S.}~\bibnamefont{Manley}},
  \bibinfo{author}{\bibfnamefont{R.~C.}~\bibnamefont{Ball}}, \bibnamefont{and}
  \bibinfo{author}{\bibfnamefont{D.~A.}~\bibnamefont{Weitz}},
  \bibinfo{journal}{Phys. Rev. Lett.} \textbf{\bibinfo{volume}{84}},
  \bibinfo{pages}{2275} (\bibinfo{year}{2000}).

\bibitem[{\citenamefont{Duri et~al.}(2009)\citenamefont{Duri, Autentieth,
  Stadler, Leupold, Chushkin, Grubel, and Gutt}}]{DuriPRL2009}
\bibinfo{author}{\bibfnamefont{A.}~\bibnamefont{Duri}},
  \bibinfo{author}{\bibfnamefont{T.}~\bibnamefont{Autenrieth}},
  \bibinfo{author}{\bibfnamefont{L.-M.} \bibnamefont{Stadler}},
  \bibinfo{author}{\bibfnamefont{O.}~\bibnamefont{Leupold}},
  \bibinfo{author}{\bibfnamefont{Y.}~\bibnamefont{Chushkin}},
  \bibinfo{author}{\bibfnamefont{G.}~\bibnamefont{Grubel}}, \bibnamefont{and}
  \bibinfo{author}{\bibfnamefont{C.}~\bibnamefont{Gutt}},
  \bibinfo{journal}{Phys. Rev. Lett.} \textbf{\bibinfo{volume}{102}},
  \bibinfo{pages}{145701} (\bibinfo{year}{2009}).

\bibitem[{\citenamefont{Bellour et~al.}(2003)\citenamefont{Bellour, Knaebel,
  Harden, Lequeux, and Munch}}]{Bellour_PRE_2003}
\bibinfo{author}{\bibfnamefont{M.}~\bibnamefont{Bellour}},
  \bibinfo{author}{\bibfnamefont{A.}~\bibnamefont{Knaebel}},
  \bibinfo{author}{\bibfnamefont{J.~L.} \bibnamefont{Harden}},
  \bibinfo{author}{\bibfnamefont{F.}~\bibnamefont{Lequeux}}, \bibnamefont{and}
  \bibinfo{author}{\bibfnamefont{J.~P.} \bibnamefont{Munch}},
  \bibinfo{journal}{Phys. Rev. E} \textbf{\bibinfo{volume}{67}},
  \bibinfo{pages}{031405} (\bibinfo{year}{2003}).

\bibitem[{\citenamefont{Bandyopadhyay et~al.}(2004)\citenamefont{Bandyopadhyay,
  Liang, Yardimci, Sessoms, Borthwick, Mochrie, Harden, and
  Leheny}}]{Bandyopadhyay_PRL_2004}
\bibinfo{author}{\bibfnamefont{R.}~\bibnamefont{Bandyopadhyay}},
  \bibinfo{author}{\bibfnamefont{D.}~\bibnamefont{Liang}},
  \bibinfo{author}{\bibfnamefont{H.}~\bibnamefont{Yardimci}},
  \bibinfo{author}{\bibfnamefont{D.~A.} \bibnamefont{Sessoms}},
  \bibinfo{author}{\bibfnamefont{M.~A.} \bibnamefont{Borthwick}},
  \bibinfo{author}{\bibfnamefont{S.~G.~J.} \bibnamefont{Mochrie}},
  \bibinfo{author}{\bibfnamefont{J.~L.} \bibnamefont{Harden}},
  \bibnamefont{and} \bibinfo{author}{\bibfnamefont{R.~L.}
  \bibnamefont{Leheny}}, \bibinfo{journal}{Phys. Rev. Lett.}
  \textbf{\bibinfo{volume}{93}}, \bibinfo{pages}{228302}
  (\bibinfo{year}{2004}).

\bibitem[{\citenamefont{Kaloun et~al.}(2005)\citenamefont{Kaloun, Skouri,
  Skouri, Munch, and Schosseler}}]{KalounPRE2005}
\bibinfo{author}{\bibfnamefont{S.}~\bibnamefont{Kaloun}},
  \bibinfo{author}{\bibfnamefont{R.}~\bibnamefont{Skouri}},
  \bibinfo{author}{\bibfnamefont{M.}~\bibnamefont{Skouri}},
  \bibinfo{author}{\bibfnamefont{J.~P.} \bibnamefont{Munch}}, \bibnamefont{and}
  \bibinfo{author}{\bibfnamefont{F.}~\bibnamefont{Schosseler}},
  \bibinfo{journal}{Phys. Rev. E} \textbf{\bibinfo{volume}{72}},
  \bibinfo{pages}{011403} (\bibinfo{year}{2005}).

\bibitem[{\citenamefont{Schosseler et~al.}(2006)\citenamefont{Schosseler,
  Kaloun, Skouri, and Munch}}]{Schoesseler_PRE_2006}
\bibinfo{author}{\bibfnamefont{F.}~\bibnamefont{Schosseler}},
  \bibinfo{author}{\bibfnamefont{S.}~\bibnamefont{Kaloun}},
  \bibinfo{author}{\bibfnamefont{M.}~\bibnamefont{Skouri}}, \bibnamefont{and}
  \bibinfo{author}{\bibfnamefont{J.~P.} \bibnamefont{Munch}},
  \bibinfo{journal}{Phys. Rev. E} \textbf{\bibinfo{volume}{73}},
  \bibinfo{pages}{021401} (\bibinfo{year}{2006}).

\bibitem[{\citenamefont{Ruta et~al.}(2012)\citenamefont{Ruta, Chushkin, Monaco,
  Cipelletti, Pineda, Bruna, Giordano, and Gonzalez-Silveira}}]{Ruta_PRL_2012}
\bibinfo{author}{\bibfnamefont{B.}~\bibnamefont{Ruta}},
  \bibinfo{author}{\bibfnamefont{Y.}~\bibnamefont{Chushkin}},
  \bibinfo{author}{\bibfnamefont{G.}~\bibnamefont{Monaco}},
  \bibinfo{author}{\bibfnamefont{L.}~\bibnamefont{Cipelletti}},
  \bibinfo{author}{\bibfnamefont{E.}~\bibnamefont{Pineda}},
  \bibinfo{author}{\bibfnamefont{P.}~\bibnamefont{Bruna}},
  \bibinfo{author}{\bibfnamefont{V.~M.}~\bibnamefont{Giordano}}, \bibnamefont{and}
  \bibinfo{author}{\bibfnamefont{M.}~\bibnamefont{Gonzalez-Silveira}},
  \bibinfo{journal}{Phys. Rev. Lett.} \textbf{\bibinfo{volume}{109}},
  \bibinfo{pages}{165701} (\bibinfo{year}{2012}).

\bibitem[{\citenamefont{Falus et~al.}(2006)\citenamefont{Falus, Borthwick,
  Narayanan, Sandy, and Mochrie}}]{FalusPRL2006}
\bibinfo{author}{\bibfnamefont{P.}~\bibnamefont{Falus}},
  \bibinfo{author}{\bibfnamefont{M.~A.}~\bibnamefont{Borthwick}},
  \bibinfo{author}{\bibfnamefont{S.}~\bibnamefont{Narayanan}},
  \bibinfo{author}{\bibfnamefont{A.~R.}~\bibnamefont{Sandy}}, \bibnamefont{and}
  \bibinfo{author}{\bibfnamefont{S.~G.~J.}~\bibnamefont{Mochrie}},
  \bibinfo{journal}{Phys. Rev. Lett.} \textbf{\bibinfo{volume}{97}},
  \bibinfo{pages}{066102} (\bibinfo{year}{2006}).

\bibitem[{\citenamefont{Narayanan et~al.}(2007)\citenamefont{Narayanan, Lee,
  Hagman, Li, and Wang}}]{NarayananPRL2007}
\bibinfo{author}{\bibfnamefont{S.}~\bibnamefont{Narayanan}},
  \bibinfo{author}{\bibfnamefont{D.~R.}~\bibnamefont{Lee}},
  \bibinfo{author}{\bibfnamefont{A.}~\bibnamefont{Hagman}},
  \bibinfo{author}{\bibfnamefont{X.}~\bibnamefont{Li}}, \bibnamefont{and}
  \bibinfo{author}{\bibfnamefont{J.}~\bibnamefont{Wang}},
  \bibinfo{journal}{Phys. Rev. Lett.} \textbf{\bibinfo{volume}{98}},
  \bibinfo{pages}{185506} (\bibinfo{year}{2007}).

\bibitem[{\citenamefont{Guo et~al.}(2009)\citenamefont{Guo, Bourret, Corbierre,
  Rucareanu, Lennox, Laaziri, Piche, Sutton, Harden, and Leheny}}]{GuoPRL2009}
\bibinfo{author}{\bibfnamefont{H.}~\bibnamefont{Guo}},
  \bibinfo{author}{\bibfnamefont{G.}~\bibnamefont{Bourret}},
  \bibinfo{author}{\bibfnamefont{M.~K.}~\bibnamefont{Corbierre}},
  \bibinfo{author}{\bibfnamefont{S.}~\bibnamefont{Rucareanu}},
  \bibinfo{author}{\bibfnamefont{R.~B.}~\bibnamefont{Lennox}},
  \bibinfo{author}{\bibfnamefont{K.}~\bibnamefont{Laaziri}},
  \bibinfo{author}{\bibfnamefont{L.}~\bibnamefont{Piche}},
  \bibinfo{author}{\bibfnamefont{M.}~\bibnamefont{Sutton}},
  \bibinfo{author}{\bibfnamefont{J.~L.}~\bibnamefont{Harden}}, \bibnamefont{and}
  \bibinfo{author}{\bibfnamefont{R.~L.}~\bibnamefont{Leheny}},
  \bibinfo{journal}{Phys. Rev. Lett.} \textbf{\bibinfo{volume}{102}},
  \bibinfo{pages}{075702} (\bibinfo{year}{2009}).

\bibitem[{\citenamefont{Caronna et~al.}(2008)\citenamefont{Caronna, Chushkin,
  Madsen, and Cupane}}]{CaronnaPRL2008}
\bibinfo{author}{\bibfnamefont{C.}~\bibnamefont{Caronna}},
  \bibinfo{author}{\bibfnamefont{Y.}~\bibnamefont{Chushkin}},
  \bibinfo{author}{\bibfnamefont{A.}~\bibnamefont{Madsen}}, \bibnamefont{and}
  \bibinfo{author}{\bibfnamefont{A.}~\bibnamefont{Cupane}},
  \bibinfo{journal}{Phys. Rev. Lett.} \textbf{\bibinfo{volume}{100}},
  \bibinfo{pages}{055702} (\bibinfo{year}{2008}).

\bibitem[{\citenamefont{Bouchaud and Pitard}(2001)}]{Bouchaud_EPJE_2001}
\bibinfo{author}{\bibfnamefont{J.-P.} \bibnamefont{Bouchaud}} \bibnamefont{and}
  \bibinfo{author}{\bibfnamefont{E.}~\bibnamefont{Pitard}},
  \bibinfo{journal}{Eur. Phys. J. E} \textbf{\bibinfo{volume}{6}},
  \bibinfo{pages}{231} (\bibinfo{year}{2001}).

\bibitem[{\citenamefont{Mourchid et~al.}(1998)\citenamefont{Mourchid, Lecolier,
  {Van Damme}, and Levitz}}]{Mourchid_Lang_1998}
\bibinfo{author}{\bibfnamefont{A.}~\bibnamefont{Mourchid}},
  \bibinfo{author}{\bibfnamefont{E.}~\bibnamefont{Lecolier}},
  \bibinfo{author}{\bibfnamefont{H.}~\bibnamefont{{Van Damme}}},
  \bibnamefont{and} \bibinfo{author}{\bibfnamefont{P.}~\bibnamefont{Levitz}},
  \bibinfo{journal}{Langmuir} \textbf{\bibinfo{volume}{14}},
  \bibinfo{pages}{4718} (\bibinfo{year}{1998}).

\bibitem[{\citenamefont{Mongondry et~al.}(2005)\citenamefont{Mongondry, Tassin,
  and Nicolai}}]{Mongondry_JCIS_2005}
\bibinfo{author}{\bibfnamefont{P.}~\bibnamefont{Mongondry}},
  \bibinfo{author}{\bibfnamefont{J.~F.} \bibnamefont{Tassin}},
  \bibnamefont{and} \bibinfo{author}{\bibfnamefont{T.}~\bibnamefont{Nicolai}},
  \bibinfo{journal}{J. Colloid Interface Sci.} \textbf{\bibinfo{volume}{283}},
  \bibinfo{pages}{397} (\bibinfo{year}{2005}).

\bibitem[{\citenamefont{Tanaka et~al.}(2005)\citenamefont{Tanaka,
  Jabbari-Farouji, Meunier, and Bonn}}]{Tanaka_PRE_2005}
\bibinfo{author}{\bibfnamefont{H.}~\bibnamefont{Tanaka}},
  \bibinfo{author}{\bibfnamefont{S.}~\bibnamefont{Jabbari-Farouji}},
  \bibinfo{author}{\bibfnamefont{J.}~\bibnamefont{Meunier}}, \bibnamefont{and}
  \bibinfo{author}{\bibfnamefont{D.}~\bibnamefont{Bonn}},
  \bibinfo{journal}{Phys. Rev. E} \textbf{\bibinfo{volume}{71}},
  \bibinfo{pages}{021402} (\bibinfo{year}{2005}).

\bibitem[{\citenamefont{{Jabbari-Farouji}
  et~al.}(2007)\citenamefont{{Jabbari-Farouji}, {Wegdam}, and
  {Bonn}}}]{Jabbari_PRL_2007}
\bibinfo{author}{\bibfnamefont{S.}~\bibnamefont{{Jabbari-Farouji}}},
  \bibinfo{author}{\bibfnamefont{G.~H.} \bibnamefont{{Wegdam}}},
  \bibnamefont{and} \bibinfo{author}{\bibfnamefont{D.}~\bibnamefont{{Bonn}}},
  \bibinfo{journal}{Phys. Rev. Lett.} \textbf{\bibinfo{volume}{99}},
  \bibinfo{pages}{065701} (\bibinfo{year}{2007}).

\bibitem[{\citenamefont{Cummins}(2007)}]{CumminsJNCS2007}
\bibinfo{author}{\bibfnamefont{H.~Z.} \bibnamefont{Cummins}},
  \bibinfo{journal}{J. Non Cryst. Sol.} \textbf{\bibinfo{volume}{353}},
  \bibinfo{pages}{3892} (\bibinfo{year}{2007}).

\bibitem[{\citenamefont{Shahin and Joshi}(2010)}]{Shahin_Langmuir_2010}
\bibinfo{author}{\bibfnamefont{A.}~\bibnamefont{Shahin}} \bibnamefont{and}
  \bibinfo{author}{\bibfnamefont{Y.~M.} \bibnamefont{Joshi}},
  \bibinfo{journal}{Langmuir} \textbf{\bibinfo{volume}{26}},
  \bibinfo{pages}{4219} (\bibinfo{year}{2010}).

\bibitem[{\citenamefont{Ruzicka and Zaccarelli}(2011)}]{RuzickaSM2011}
\bibinfo{author}{\bibfnamefont{B.}~\bibnamefont{Ruzicka}} \bibnamefont{and}
  \bibinfo{author}{\bibfnamefont{E.}~\bibnamefont{Zaccarelli}},
  \bibinfo{journal}{Soft Matter} \textbf{\bibinfo{volume}{7}},
  \bibinfo{pages}{1268} (\bibinfo{year}{2011}).

\bibitem[{\citenamefont{Tudisca et~al.}(2012)\citenamefont{Tudisca, Ricci,
  Angelini, and Ruzicka}}]{TudiscaRSC2012}
\bibinfo{author}{\bibfnamefont{V.}~\bibnamefont{Tudisca}},
  \bibinfo{author}{\bibfnamefont{M.}~\bibnamefont{Ricci}},
  \bibinfo{author}{\bibfnamefont{R.}~\bibnamefont{Angelini}}, \bibnamefont{and}
  \bibinfo{author}{\bibfnamefont{B.}~\bibnamefont{Ruzicka}},
  \bibinfo{journal}{RSC Advances} \textbf{\bibinfo{volume}{2}},
  \bibinfo{pages}{11111} (\bibinfo{year}{2012}).

\bibitem[{\citenamefont{Ruzicka et~al.}(2010)\citenamefont{Ruzicka, Zulian,
  Zaccarelli, Angelini, Sztucki, Moussa\"\i{}d, and Ruocco}}]{RuzickaPRL2010}
\bibinfo{author}{\bibfnamefont{B.}~\bibnamefont{Ruzicka}},
  \bibinfo{author}{\bibfnamefont{L.}~\bibnamefont{Zulian}},
  \bibinfo{author}{\bibfnamefont{E.}~\bibnamefont{Zaccarelli}},
  \bibinfo{author}{\bibfnamefont{R.}~\bibnamefont{Angelini}},
  \bibinfo{author}{\bibfnamefont{M.}~\bibnamefont{Sztucki}},
  \bibinfo{author}{\bibfnamefont{A.}~\bibnamefont{Moussa\"\i{}d}},
  \bibnamefont{and} \bibinfo{author}{\bibfnamefont{G.}~\bibnamefont{Ruocco}},
  \bibinfo{journal}{Phys. Rev. Lett.} \textbf{\bibinfo{volume}{104}},
  \bibinfo{pages}{085701} (\bibinfo{year}{2010}).

\bibitem[{\citenamefont{{Beck} et~al.}(1999)\citenamefont{{Beck}, {H\"artl},
  and {Hempelmann}}}]{BeckJCP1999}
\bibinfo{author}{\bibfnamefont{C.}~\bibnamefont{{Beck}}},
  \bibinfo{author}{\bibfnamefont{W.}~\bibnamefont{{H\"artl}}},
  \bibnamefont{and}
  \bibinfo{author}{\bibfnamefont{R.}~\bibnamefont{{Hempelmann}}},
  \bibinfo{journal}{J. Chem.Phys} \textbf{\bibinfo{volume}{111}},
  \bibinfo{pages}{8209} (\bibinfo{year}{1999}).

\bibitem[{\citenamefont{Zaccarelli et~al.}(2008)\citenamefont{Zaccarelli,
  Andreev, Sciortino, and Reichman}}]{ZacPRL2008}
\bibinfo{author}{\bibfnamefont{E.}~\bibnamefont{Zaccarelli}},
  \bibinfo{author}{\bibfnamefont{S.}~\bibnamefont{Andreev}},
  \bibinfo{author}{\bibfnamefont{F.}~\bibnamefont{Sciortino}},
  \bibnamefont{and} \bibinfo{author}{\bibfnamefont{D.~R.}
  \bibnamefont{Reichman}}, \bibinfo{journal}{Phys. Rev. Lett.}
  \textbf{\bibinfo{volume}{100}}, \bibinfo{pages}{195701}
  (\bibinfo{year}{2008}).

\bibitem[{\citenamefont{Kang et~al.}(2013)\citenamefont{Kang, Kirkpatrick, and
  Thirumalai}}]{KangPRE2013}
\bibinfo{author}{\bibfnamefont{H.}~\bibnamefont{Kang}},
  \bibinfo{author}{\bibfnamefont{T.~R.} \bibnamefont{Kirkpatrick}},
  \bibnamefont{and}
  \bibinfo{author}{\bibfnamefont{D.}~\bibnamefont{Thirumalai}},
  \bibinfo{journal}{Phys. Rev. E} \textbf{\bibinfo{volume}{88}},
  \bibinfo{pages}{042308} (\bibinfo{year}{2013}).

\bibitem[{\citenamefont{Madsen et~al.}(2010)\citenamefont{Madsen, Leheny, Guo,
  Sprung, and Czakkel}}]{MadsenNJP2010}
\bibinfo{author}{\bibfnamefont{A.}~\bibnamefont{Madsen}},
  \bibinfo{author}{\bibfnamefont{R.}~\bibnamefont{Leheny}},
  \bibinfo{author}{\bibfnamefont{H.}~\bibnamefont{Guo}},
  \bibinfo{author}{\bibfnamefont{M.}~\bibnamefont{Sprung}}, \bibnamefont{and}
  \bibinfo{author}{\bibfnamefont{O.}~\bibnamefont{Czakkel}},
  \bibinfo{journal}{New J. of Phys.} \textbf{\bibinfo{volume}{12}},
  \bibinfo{pages}{055001} (\bibinfo{year}{2010}).

\bibitem[{\citenamefont{{Ruzicka} et~al.}(2004)\citenamefont{{Ruzicka},
  {Zulian}, and {Ruocco}}}]{RuzickaPRL2004}
\bibinfo{author}{\bibfnamefont{B.}~\bibnamefont{{Ruzicka}}},
  \bibinfo{author}{\bibfnamefont{L.}~\bibnamefont{{Zulian}}}, \bibnamefont{and}
  \bibinfo{author}{\bibfnamefont{G.}~\bibnamefont{{Ruocco}}},
  \bibinfo{journal}{Phys. Rev. Lett.} \textbf{\bibinfo{volume}{93}},
  \bibinfo{pages}{258301} (\bibinfo{year}{2004}).

\bibitem[{\citenamefont{Angelini et~al.}(2013)\citenamefont{Angelini, Zulian,
  Fluerasu, Madsen, Ruocco, and Ruzicka}}]{AngeliniSM2013}
\bibinfo{author}{\bibfnamefont{R.}~\bibnamefont{Angelini}},
  \bibinfo{author}{\bibfnamefont{L.}~\bibnamefont{Zulian}},
  \bibinfo{author}{\bibfnamefont{A.}~\bibnamefont{Fluerasu}},
  \bibinfo{author}{\bibfnamefont{A.}~\bibnamefont{Madsen}},
  \bibinfo{author}{\bibfnamefont{G.}~\bibnamefont{Ruocco}}, \bibnamefont{and}
  \bibinfo{author}{\bibfnamefont{B.}~\bibnamefont{Ruzicka}},
  \bibinfo{journal}{Soft Matter} \textbf{\bibinfo{volume}{9}},
  \bibinfo{pages}{10955} (\bibinfo{year}{2013}).

\bibitem[{\citenamefont{Saw et~al.}(2009)\citenamefont{Saw, Ellegaard, Kob, and
  Sastry}}]{SawPRL2009}
\bibinfo{author}{\bibfnamefont{S.}~\bibnamefont{Saw}},
  \bibinfo{author}{\bibfnamefont{N.~L.} \bibnamefont{Ellegaard}},
  \bibinfo{author}{\bibfnamefont{W.}~\bibnamefont{Kob}}, \bibnamefont{and}
  \bibinfo{author}{\bibfnamefont{S.}~\bibnamefont{Sastry}},
  \bibinfo{journal}{Physical Review Letters} \textbf{\bibinfo{volume}{103}},
  \bibinfo{pages}{248305} (\bibinfo{year}{2009}).

\bibitem[{\citenamefont{{El Masri} et~al.}(2010)\citenamefont{{El Masri},
  Berthier, and Cipelletti}}]{MasriPRE2010}
\bibinfo{author}{\bibfnamefont{D.}~\bibnamefont{{El Masri}}},
  \bibinfo{author}{\bibfnamefont{L.}~\bibnamefont{Berthier}}, \bibnamefont{and}
  \bibinfo{author}{\bibfnamefont{L.}~\bibnamefont{Cipelletti}},
  \bibinfo{journal}{Phys. Rev. E} \textbf{\bibinfo{volume}{82}},
  \bibinfo{pages}{031503} (\bibinfo{year}{2010}).

\end{thebibliography}

}

\end{document}